\documentclass[prb,twocolumn,showpacs]{revtex4}
\usepackage[latin1]{inputenc}
\usepackage{graphicx}
\usepackage{amssymb}
\usepackage{amsmath}
\usepackage{xspace}
\usepackage{citesort}
\usepackage{dcolumn}
\usepackage{bm}

\newcommand{\etal}[0]{{\it et al.}\@\xspace}
\newcommand{\ie}[0]{i.e.\@\xspace}
\newcommand{\eg}[0]{e.g.\@\xspace}
\newcommand{\omb}[0]{\overline{\omega}}
\newcommand{\om}[0]{\omega}
\newcommand{\si}[0]{\sigma}
\newcommand{\en}[0]{\epsilon}
\renewcommand{\sf}[1]{A_{#1}(k,\en)}
\newcommand{\las}[0]{\langle}
\newcommand{\ras}[0]{\rangle}
\newcommand{\la}[0]{\left\las}
\newcommand{\ra}[0]{\right\ras}
\newcommand{\ket}[1]{\left|#1\ra}  
\newcommand{\bra}[1]{\la#1\right|} 
\newcommand{\rmi}{\mathrm{i}}
\renewcommand{\Im}[0]{\text{Im}}
\newcommand{\G}[0]{\mathcal{G}}
\newcommand{\Ub}[0]{\overline{U}}
\newcommand{\rD}[0]{\text{D}}
\newcommand{\bk}{\bm{k}}
\newcommand{\UP}{\uparrow}
\newcommand{\DO}{\downarrow}
\newcommand{\Ep}{E_\text{P}}
\newcommand{\Nph}{N_\text{ph}}
\newcommand{\nag}{\phantom{\dag}}

\begin{document}


\title{Single-particle spectral function of the Holstein-Hubbard bipolaron}

\author{Martin Hohenadler}
\email{hohenadler@itp.tugraz.at}
\author{Markus Aichhorn}\author{Wolfgang \surname{von der Linden}}
\affiliation{%
  Institute for Theoretical and Computational Physics, Graz University of
  Technology, Petersgasse 16, 8010 Graz, Austria}

\begin{abstract}
  The one-electron spectral function of the Holstein-Hubbard bipolaron in one
  dimension is studied using cluster perturbation theory together with the
  Lanczos method. In contrast to other approaches, this allows one to
  calculate the spectrum at continuous wavevectors and thereby to
  investigate, for the first time, the dispersion and the spectral weight of
  quasiparticle features. The formation of polarons and bipolarons, and their
  manifestation in the spectral properties of the system, is studied for the
  cases of intermediate and large phonon frequencies, with and without
  Coulomb repulsion. A good agreement is found with the most accurate
  calculations of the bipolaron band dispersion available. Pronounced
  deviations of the bipolaron band structure from a simple tight-binding band
  are observed, which can be attributed to next-nearest-neighbor hopping
  processes.
\end{abstract}


\maketitle

\section{\label{sec:introduction}Introduction}

In recent years, angle-resolved photoemission spectroscopy (ARPES) has proved
to be very helpful in obtaining information about the electronic states of
strongly correlated systems. While a lot of data is available from
experiments, reliable theoretical calculations of the one-particle spectral
function---which can often be regarded as being proportional to the ARPES
spectrum---for popular models of, \eg, the Hubbard or $t$-$J$ type, are
usually very demanding.  As a consequence, many interesting problems of
condensed matter physics have not been investigated systematically regarding
their spectral properties in a satisfactory way. Among them is the ``{\it
  bipolaron problem}'' of two electrons, which can form a bound state even in
the presence of strong Coulomb repulsion if they are coupled to phonons.
Despite the long history of this problem, bipolaron formation is still
subject of ongoing discussion due to its potential role, \eg, in
high-temperature superconductors\cite{BYMdLBi92} and
manganites,\cite{David_AiP} two classes of materials studied extensively over
the past decade.

Most existing results for the Holstein-Hubbard (HH) model considered here
have been obtained using exact diagonalization (ED).  Apart from a systematic
error due to the necessary truncation of the Hilbert space, this method gives
exact results for the one-electron spectrum, but is restricted to rather
small systems, especially for small phonon frequencies and/or strong
electron-phonon coupling. Consequently, it is difficult to study the
dispersion of the spectral peaks throughout the Brillouin zone. To overcome
this limitation, we employ here cluster perturbation theory (CPT), extending
the recent application to the Holstein model with one
electron.\cite{HoAivdL03} In contrast to ED, CPT permits to calculate the
spectral function for continuous wavevectors. Moreover, finite-size effects
are strongly reduced compared to direct diagonalization of small clusters,
and results are much more realistic than previous work based on, \eg, a
two-site system.\cite{RaTh92,dMeRa97,dMeRa98} CPT becomes exact in the weak-
and strong-coupling regime, and has been successfully applied also to other
problems.\cite{SePePL00,SePePl02,ZaEdArHa00,ZaEdArHa02,DaArHa02,SeTr03,DaAiHaArPo04,AiEvvdLPo04}
A review of cluster methods for strongly correlated systems is by Maier
\etal\cite{MaJaPrHe04} Here we would merely like to point out the recent
application of the dynamical cluster approximation to the half-filled
Holstein model by Hague.\cite{Ha03}

In this work, we study in detail the formation of polarons and bipolarons,
and its dependence on phonon frequency, electron-phonon and electron-electron
interaction. To test the reliability of our results, and to interpret the
quasiparticle (QP) features in the spectra, comparison is made to ED, as well
as to the most accurate approach to the one-dimensional HH bipolaron
currently available, namely the variational diagonalization
method.\cite{BoKaTr00} Moreover, we also investigate the form of the
bipolaron band dispersion and compare it to that of models with
nearest-neighbor- and next-nearest-neighbor hopping.

This paper is organized as follows. We begin with a review of the HH
bipolaron in Sec.~\ref{sec:holstein}.  In Sec.~\ref{sec:method}, we discuss
some details of the application of CPT.  Results are presented in
Sec.~\ref{sec:results} and, finally, Sec.~\ref{sec:summary} contains our
conclusions.

\section{\label{sec:holstein}The Holstein-Hubbard model}

The HH model is defined by the Hamiltonian
\begin{eqnarray}\label{eq:holstein}\nonumber
  H
  &=&
  -t\sum_{\las ij\ras\si} \left(c^\dag_{i\si} c^{\phantom{\dag}}_{j\si} + \text{h.c.}\right)
  +\om\sum_i  b^{\dag}_i b^{\phantom{\dag}}_i\\
  && -g\sum_i n_i (b^\dag_i + b^{\phantom{\dag}}_i)
  + U \sum_i n_{i\UP} n_{i\DO}\,.
\end{eqnarray}
Here $c^\dag_{i\sigma}$ ($c^{\nag}_{i\sigma}$) and $b^{\dag}_i$
($b^{\nag}_i$) create (annihilate) an electron of spin $\si$, and a phonon of
energy $\om$ ($\hbar=1$) at site $i$, respectively, and $n_i=\sum_\si
n_{i\si}$ with $n_{i\si}=c^\dag_{i\si} c^{\phantom{\dag}}_{i\si}$. The first
two terms correspond to the kinetic energy of the electrons and the kinetic
and elastic energy of the phonons, respectively. The electron-phonon (el-ph)
and electron-electron (el-el) interaction is described by the third and forth
term. We have three model parameters, namely the amplitude for
nearest-neighbor hopping $t$, the phonon frequency $\om$, the el-ph coupling
constant $g$, and the el-el interaction strength $U>0$. For $U=0$,
Eq.~(\ref{eq:holstein}) is identical to the pure Holstein model, while for
$g=0$ we recover the Hubbard model. We introduce the commonly used
dimensionless coupling constant $\lambda =2g^2/(\om W)$, where $W=4t\rD$ is
the bare bandwidth in D dimensions. We further define the dimensionless
parameters $\omb=\om/t$ and $\Ub=U/t$, and express all energies in units of
$t$. Consequently, the independent parameters of the model are $\omb$, $\Ub$,
and $\lambda$. In the sequel, we shall also use the polaron binding energy
$\Ep=\lambda W/2$, which emerges as a natural parameter from the Lang-Firsov
transformation.\cite{LangFirsov} Finally, the lattice constant is taken to be
unity.

Owing to the complexity of even the two-electron problem, we will not discuss
effects of bipolaron-bipolaron interaction here. An investigation of the
latter, which will definitely play an important role in real materials,
requires a study of the HH model with many electrons (see, \eg,
Ref.~\onlinecite{FeWeHaWeBi03} and references therein), which is beyond the
scope of our method in its present form.

There exists a considerable amount of work on the HH bipolaron, although it
is by far not as well understood as the simpler one-electron case. In the
sequel, we restrict our discussion to new developments in the field. A very
complete review of earlier work has been given by Alexandrov and
Mott.\cite{AlMo95}

While the pairing of electrons in momentum space can be accurately described
by Migdal-Eliashberg theory\cite{Ma90} for weak-enough coupling, no reliable
theory is available for the formation of bipolarons---corresponding to
pairing of electrons in real space---at intermediate to strong el-ph
interaction. In recent years, progress was made using either variational
approaches\cite{CoEcSo84,EmYeBe92,LaMaPu97,PrAu98,PrAu99,PrAu00,deFiCaIaMaPeVe01}
or, more importantly, unbiased numerical studies based on
ED,\cite{RaTh92,Marsiglio95,WeRoFe96,dMeRa97,dMeRa98} variational
diagonalization,\cite{FeRoWeMi95,WeFeWeBi00,BoKaTr00,ElShBoKuTr03}
DMRG,\cite{ZhJeWh99} and QMC.\cite{deRaLa86,Mac04,HovdL04} The ED and DMRG
calculations were restricted to rather small systems consisting of
two,\cite{RaTh92,dMeRa97,dMeRa98} four,\cite{Marsiglio95} six,\cite{ZhJeWh99}
eight\cite{FeRoWeMi95,WeRoFe96} or twelve sites,\cite{WeFeWeBi00} while the
methods of Refs.~\onlinecite{BoKaTr00},~\onlinecite{deRaLa86}
and~\onlinecite{Mac04} are almost free of finite-size effects. The larger
number of phonon states required to obtain converged results makes numerical
studies with ED methods even more challenging than for a single electron,
especially for small phonon frequencies.

Since the HH model represents a simplified description of the situation in
real materials, it is highly desirable to study more complex models. To this
end, it is interesting to note that the QMC methods of de Raedt and
Lagendijk\cite{deRaLa86} and Macridin \etal\cite{Mac04} may be generalized to
include dispersive phonons.  Furthermore, both approaches can be applied to
models with long-range Coulomb interaction,\cite{deRaLa86,Mac04} similar to
the work of Bon\v{c}a and Trugman.\cite{BoTr01} Finally, bipolaron formation
in a model with Jahn-Teller modes---as present, \eg, in perovskite
manganites---has recently been investigated by Shawish
\etal\cite{ElShBoKuTr03}

To discuss bipolaron formation in the HH model, we have to distinguish
between two cases. The two electrons can either have the same or opposite
spin, which leads to a singlet or triplet state, respectively. We consider
these possibilities separately.

\subsubsection{Singlet state}

For two electrons in a singlet state, the formation of a bound bipolaron
state in the absence of Coulomb interaction originates from the fact that the
potential well---arising from a displacement of the oscillators---around an
occupied lattice site deepens in the presence of a second electron. This may
easily be seen in the atomic limit $t=0$, using the Lang-Firsov
transformation.\cite{LangFirsov} On different lattice sites, each electron
gains an energy $-\Ep$ by distorting the lattice, whereas the energy shift
becomes $-4\Ep$ if both particles occupy the same site ({\it small} or {\it
  onsite bipolaron}). For $t\neq0$, the competition between the kinetic
energy of the electrons on the one hand and the displacement or lattice
energy on the other hand determines the cross over from a state with two
weakly bound polarons, sometimes also referred to as a {\it large bipolaron},
for $\lambda<\lambda_\text{c}$ to a small bipolaron for
$\lambda>\lambda_\text{c}$, where $\lambda_\text{c}$ denotes the critical
value of the el-ph coupling.  In Sec.~\ref{sec:holstein}, $\lambda$ has been
defined as $\lambda=2\Ep/W$, \ie, as the ratio of the energy gain due to
polaron formation to the kinetic energy of a free electron. While
$\lambda_\text{c}=1$ in the adiabatic regime for the small-polaron cross over
in the model with one electron (see, \eg, discussion in
Ref.~\onlinecite{HoEvvdL03}), here we expect $\lambda_\text{c}=0.5$ (for
$\omb\ll1$) due to the energy gain of $-2\Ep$ per electron compared to $-\Ep$
in the single polaron problem. This is well confirmed by the calculations of
Wellein \etal,\cite{WeRoFe96} who find a strong decrease of the kinetic
energy near $\lambda=0.5$ for $\omb=0.4$.

For $\omb\gg1$, the lattice energy becomes important since the trapping of
the carriers requires a sizable lattice distortion. This gives rise to the
additional condition $2\sqrt{\Ep/\om}>1$ for a small
bipolaron.\cite{ChRaFe98} Similar to the one-electron problem, the cross over
is very gradual in the nonadiabatic regime.\cite{WeRoFe96} The correlation or
binding of the two electrons depends crucially on the phonon frequency, since
the latter determines the maximum distance across which the two particles
feel an attractive interaction due to the phonons. Up to second order in $g$,
this coupling is given by
\begin{equation}\label{eq:veff_pt}
  U_\text{eff}(\en)
  =
  g^2 \mathcal{D}_\text{ph}(\bm{q},\en)
  =
  -\frac{2g^2\om}{\om^2 - \en^2}\,,
\end{equation}
where $\mathcal{D}_\text{ph}(\bm{q},\en)$ denotes the phonon propagator.
Equation~(\ref{eq:veff_pt}) reveals that the energy-dependent interaction is
attractive for $\en<\om$, and becomes instantaneous in the antiadiabatic limit
$\om\rightarrow\infty$ where $U_\text{eff}=-2g^2/\om$.  Hence, the binding
always decreases with increasing phonon frequency.\cite{Marsiglio95}

For $U>0$, there is a competition between the retarded, attractive
interaction mediated by the phonons and the instantaneous, repulsive Hubbard
interaction. Consequently, a state with two unbound polarons---stabilized by
the onsite repulsion---can exist for sufficiently weak el-ph
coupling.\cite{WeRoFe96} This is in contrast to the extended HH model with
long-range interaction, in which a bipolaron state is formed irrespective of
the value of $U$.\cite{WeWeFe02} The effective el-el interaction in the HH
model determining the nature of the bipolaron state is
\begin{equation}\label{eq:Ueff}
  U_\text{eff}
  =
  U-2\Ep\,.
\end{equation}
From this result, which can be obtained either from the generalization of
Eq.~(\ref{eq:veff_pt}) to $U\neq0$ in the limit $\om\rightarrow\infty$, or in
the antiadiabatic strong-coupling limit,\cite{Mac04} one may be tempted to
expect a bipolaron state to exist only for $U_\text{eff}<0$, \ie, if there is
a net attractive interaction between the particles. While this is true for
the effective Hubbard model onto which the HH model maps in the antiadiabatic
strong-coupling limit, a consideration of virtual hopping processes leads to
the less stringent condition $U<4\Ep$.\cite{BoKaTr00} The energy gain due to
virtual exchange processes of two electrons on neighboring lattice
sites---not suppressed by strong el-ph coupling---permits the formation of a
weakly bound {\it intersite bipolaron} with the two electrons most likely to
reside on neighboring lattice sites.\cite{BoKaTr00,Mac04} A phase diagram for
bipolaron formation as a function of $\lambda$ and $\omb$ in one dimension
has been presented by Wei{\ss}e \etal\cite{WeFeWeBi00} Eventually, for
sufficiently strong el-ph coupling $2\Ep\gtrsim U$, the effective onsite
potential $U_\text{eff}$ becomes attractive, and a small bipolaron is formed.

Starting from a small bipolaron, a cross over to an intersite bipolaron takes
place when the Coulomb interaction becomes large
enough.\cite{BoKaTr00,WeWeFe02,Mac04} The intersite bipolaron has a much
smaller effective mass than the small bipolaron and may therefore also exist
as a mobile carrier in real materials.\cite{BoKaTr00} In the adiabatic limit
$\omb=0$, the onsite--intersite bipolaron transition has been shown to be of
first order,\cite{PrAu98,PrAu99} but for finite phonon frequencies it is
expected to happen in a more gradual way because of retardation effects, in
agreement with recent calculations.\cite{BoKaTr00} Estimates for the region
of existence of the intersite bipolaron state for $\omb=1$ are $U<2\Ep$ for
weak coupling, and $U<4\Ep$ for strong el-ph coupling,\cite{BoKaTr00} and
phase diagrams in the $(U,\lambda)$-plane have been reported in
one\cite{BoKaTr00} and two dimensions.\cite{Mac04} While the above conditions
are quite accurate in the nonadiabatic regime $\omb\geq1$, the case
$\omb\ll1$ remains an open problem.

Finally, the physically most interesting regime, which is unfortunately also
the most difficult case to treat theoretically, is defined by $\omb\ll1$, and
a Coulomb repulsion at least as large as the attractive interaction due to
the el-ph coupling.

\subsubsection{Triplet state}

For two electrons of the same spin, the Pauli principle forbids double
occupation of a site. In principle, a bound state may be formed with the two
particles being located on different lattice sites. While two electrons can
lower their energy by sharing a lattice distortion in the large bipolaron
regime, especially for small phonon frequencies, the exchange process
stabilizing the singlet intersite bipolaron state at intermediate-to-strong
el-ph coupling and $U>0$ is not strong enough to bind two polarons in a
triplet intersite state.\cite{BoKaTr00} Furthermore, for $U<\infty$, the
ground-state energy of the triplet state is always larger than for a singlet
state because two particles with parallel spin cannot occupy the same
$\bm{k}=0$ energy level. Finally, the singlet and triplet states become
degenerate in the limit $U\rightarrow\infty$.

\section{\label{sec:method}Method}

As mentioned above, here we use CPT in combination with the Lanczos recursion
method.\cite{CuWi85} Details about the application to el-ph problems have
been given in Ref.~\onlinecite{HoAivdL03}, henceforth also referred to as I.
The major difficulty we are facing in the present case is the larger number
of phonon states needed to obtain converged results.  From a physical point
of view, this is not surprising since each of the two electrons will create a
lattice distortion or phonon cloud, whereas there is only one dressed
particle (polaron) in the case of the Holstein polaron considered in I.
However, in addition to the simple doubling of the number of particles, it
has been shown by previous authors\cite{Marsiglio95,WeRoFe96,ZhJeWh99} that
multiphonon states play a more important role for the bipolaron as a result
of the phonon-mediated binding.

ED (and also CPT) for el-ph systems is affected both by finite-size effects
and the truncation error due to the restricted number of phonon states kept
in calculations. Obviously, if one used very small clusters, good convergence
with respect to the phonons could be achieved even for strong el-ph coupling.
On the other hand, for small numbers of phonon states, rather large clusters
can be studied. The approach which has been widely used in the past is to
require the truncation error, \eg, of the ground-state energy, to be smaller
than a certain limit, and to use the maximal cluster size which can be
handled for this number of phonons.  Here, an additional challenge arises
form the fact that the diagonalization of the cluster has to be performed for
open boundary conditions.\cite{HoAivdL03} Consequently, one cannot exploit
translational symmetry to reduce the dimensionality of the Hilbert space.
However, this drawback is clearly outweighed by the advantages of CPT
outlined below.

We would like to briefly discuss some interesting features of CPT. The method
is based on a breakup of the infinite lattice into clusters of $N$ sites,
say.\cite{SePePl02} The one-electron cluster Green function, denoted here as
$G_{ab,\si}$ [see Eq.~(4) in I], of the model under consideration is
calculated for one of these (identical) clusters using open boundary
conditions. This can be done, \eg, using ED or analytical
approaches.\cite{HoAivdL03} The hopping between adjacent clusters is then
treated as a perturbation to obtain the Green function of the original
system, $\G_\si(\bm{k},\en)$.\cite{SePePl02} A basic limitation of the theory
is that the Hamiltonian must not contain any nonlocal interactions, except
for one-electron terms. Additionally, CPT in its present form can only be
used to calculate one-particle Green functions.\cite{SePePl02} Therefore,
interesting observables such as, \eg, el-el correlation functions or
transport properties are not yet available.

From the nature of the approximation made, it is clear that CPT will work
particularly well if the local interactions dominate the physics of the
system, \ie, for the case of the HH model (\ref{eq:holstein}), if $g, U \gg
t$. This point will be illustrated in Sec.~\ref{sec:results}. The method
becomes exact in the atomic limit $t=0$, for noninteracting electrons
($g,U=0$), as well as for $N=\infty$.\cite{SePePl02} The quality of the
results obtained with CPT has been tested for several
models,\cite{SePePL00,SePePl02,ZaEdArHa00,ZaEdArHa02,DaArHa02,SeTr03,DaAiHaArPo04,AiEvvdLPo04}
and a very good agreement with other work has been found.  In I, we pointed
out the occurrence of finite-size effects which show up as additional peaks
in the corresponding one-electron spectral function. The weight of the latter
reduces quickly as $N\rightarrow\infty$, so that the spectrum is not affected
significantly.

One of the most important advantages of CPT is that it gives, in principle,
results for an infinite system, although the approximate treatment of the
intercluster hopping introduces some finite-size effects, which can be
systematically reduced by increasing the cluster size. As a consequence, the
one-electron Green function can be calculated for any wavevector in the
Brillouin zone, even for $N=1$. This allows one to study the dispersion of
the QP peaks, in strong contrast to standard ED methods on finite clusters,
for which only a few points in momentum space are accessible, owing to the
rather small values of $N\approx2$\,--\,20 usually used.

Since we only consider the one-dimensional HH model in the sequel, we shall
adopt the notation accordingly. We are interested in the one-electron Green
function
\begin{equation}\label{eq:G_lanczos}
  \G_{\si}(k,\en)
  =
  \bra{\DO}
  c^{\nag}_{k\si}
  \frac{1}{\en-H}
  c^\dag_{k\si}
  \ket{\DO}
  \,,
\end{equation}
where $\ket{\DO}$ denotes the ground state with one electron of spin down,
and $\si=\UP,\DO$. Equation~(\ref{eq:G_lanczos}) only contains the inverse
photoemission part of the total one-electron Green function. In the case of
$\G_{\UP}$, the second part---corresponding to photoemission (PE)---vanishes,
since there is no $\UP$ electron in the ground state. Moreover, for
$\G_{\DO}$, PE is identical to the spectrum of a single polaron, which has
been studied in detail in I. The situation would be different if we started
with a two-electron (singlet or triplet) ground state. Then, the PE part of
the one-electron spectral function also contains valuable information.
However, due to limited computer memory, such computations involving
three-electron states are not possible with the code used here.

In Eq.~(\ref{eq:G_lanczos}), we have omitted the energy $E^\DO_0$ of the
ground state $\ket{\DO}$, which usually enters in the form $H-E^\DO_0$, to
permit direct comparison with the singlet bipolaron band dispersion
$E^{\UP\DO}(k)$ in Sec.~\ref{sec:results}.  The one-electron spectral
function is related to the Green function~(\ref{eq:G_lanczos}) via
\begin{equation}\label{eq:Akw}
  \sf{\si}
  =
  -\pi^{-1}\lim_{\eta\rightarrow 0^+}\Im\,\G_{\si}(k,\en+\rmi\eta)\,.
\end{equation}
To calculate the cluster Green function by ED, a truncation of the phonon
Hilbert space is necessary, and we use the same truncation scheme as in I.
The number of phonon states $\Nph$ will be chosen so as to push the
truncation error
$\Delta\equiv|E^{\UP\DO}_0(\Nph+1)-E^{\UP\DO}_0(\Nph)|/|E^{\UP\DO}_0(\Nph)|$
of the energy $E^{\UP\DO}_0$ of the two-electron ground state $\ket{\UP\DO}$
below $10^{-4}$.  The use of $E^{\UP\DO}_0$ to monitor convergence with
respect to $\Nph$ comes from the observation that---for the same number of
phonons---the truncation error of the latter is always smaller than for the
triplet state $\ket{\DO\DO}$. This may be ascribed to the fact that for two
electrons of the same spin, no bound onsite or intersite state exists. In particular, there will
be no large local lattice distortion surrounding an onsite bipolaron, the
description of which requires a significant number of phonons. In previous
work on the HH bipolaron, using ED with periodic boundary
conditions,\cite{RaTh92,Marsiglio95,WeRoFe96,dMeRa97,dMeRa98} the truncation
error was usually smaller than $10^{-6}$. However, these methods were
restricted to only a few $k$ vectors. Furthermore, our calculations show that
even a relative error $\Delta=10^{-4}$ ensures satisfactory convergence of
the one-electron spectrum. The smaller number of phonon states enables us to
use larger clusters and thereby significantly diminish finite-size effects
since, within the CPT, even an increase $N\rightarrow N+1$ noticeably
improves the results. Once the cluster size has been fixed, we use the
maximal possible number of phonons. The accuracy $\Delta$ varies for the
different calculations and will be reported in each figure.

In its present form, our method is restricted to the nonadiabatic regime
$\omb\geq1$, except for weak el-ph coupling. To study smaller phonon
frequencies---relevant to, \eg, to transition metal oxides---a combination
with variational diagonalization techniques, or the use of shared-memory
systems would be necessary. As in I, we restrict our calculations to the
spectral function, which is the most fundamental quantity that can be
obtained from CPT.\cite{SePePl02}

\section{\label{sec:results}Results}

The one-electron spectral function of the problem considered here has been
calculated before using ED\cite{RaTh92,dMeRa97,dMeRa98} and
DMRG,\cite{ZhJeWh99} both in one dimension. However, results were only given
for $k=0$, and for very small systems with $N=2$ and $N=6$, respectively.
With the above methods, and for periodic boundary conditions, the spectral
function~(\ref{eq:Akw}) can be evaluated for $N$ different wavevectors, of
which only $N/2+1$ are physically nonequivalent. This makes it difficult or
even impossible to study the dispersion of QP features.  Recently, a
parallelized DMRG code has been developed,\cite{HaJeFeWe03} which allows
studies of one-dimensional Holstein models on very large clusters even at
half filling.\cite{FeWeHaWeBi03} However, the calculation of spectral
functions within DMRG is very time-consuming, since it has to be done
separately for each point on the energy axis.  Several authors have also
calculated dressed spectral functions,\cite{dMeRa97,dMeRa98,ZhJeWh99} with
the fermion operators in Eq.~(\ref{eq:G_lanczos}) replaced by their
Lang-Firsov transformed (\ie, dressed) counterparts, as well as pair spectral
functions.\cite{ZhJeWh99} The corresponding spectra show a simplified
structure in certain regimes, indicating that polarons and bipolarons are
``good'' QP's for these parameters.

De Mello and Ranninger\cite{dMeRa97} have pointed out that to study the cross
over between polarons and bipolarons it is, in general, necessary to
investigate both, photoemission and inverse photoemission. This can easily be
understood by considering electron emission from the two-electron singlet
ground state, \ie, the Green function
$\bra{\UP\DO}c^{\dag}_{k\UP}(\en-H)^{-1}c^{\nag}_{k\UP}\ket{\UP\DO}$.
Depending on the parameters, $\ket{\UP\DO}$ may either consist of two weakly
bound polarons or a bipolaron. Consequently, photoemission spectra will only
show a single QP band. In contrast, the Green function~(\ref{eq:G_lanczos})
with $\si=\UP$ corresponds to adding an $\UP$ electron to the one-electron
ground state $\ket{\DO}$. For example, the additional particle can either go
into the ground state to form a bipolaron, or into an excited polaron state.
In general, we therefore expect two QP bands in the spectral function, whose
weights, positions and widths vary with $\omb$, $\Ub$ and $\lambda$.

As we will compare our findings with the variational diagonalization method
(VDM) of Bon\v{c}a \etal,\cite{BoKaTr00,ElShBoKuTr03} we would like to
comment on the accuracy of the latter. The problem is defined on an infinite
system, so that the approach is free of boundary finite-size effects.
However, the method involves a variationally determined Hilbert space with
two variational parameters, namely the maximal allowed distance between
electrons and phonons, and between the two electrons, respectively. For the
bipolaron problem under consideration, the limiting parameter in the regime
$\omb\geq1$ is the maximum distance $N_\text{h}$ between the two electrons.
The results presented here have been obtained using $N_\text{h}\leq18$. While
the method gives very accurate results---with errors smaller than the
linewidth in the figures---for the case of a small bipolaron ($U\ll2\Ep$), it
is less reliable (relative errors $\lesssim 1\%$) for strong onsite repulsion
$U\gg2\Ep$ favoring two weakly bound polarons, similar to ED and CPT.  Due to
additional towers of phonon excitations that are located in the neighborhood
of the electron sites, the method achieves good convergence in the small
bipolaron regime even for strong coupling. Nevertheless, the adiabatic regime
$\omb\ll1$ represents a difficult problem, as is the case for other
approaches. Finally, as in CPT, results may be obtained at any wavevector.

We shall see below that there is a close correspondence of the QP bands in
the spectra to the polaron and bipolaron dispersion relations denoted here as
$E^\UP(k)$ and $E^{\UP\DO}(k)$, respectively. The notation $E^\UP(k)$ is
convenient, but there is no spin-dependence in the case of a single electron,
\ie, $E^\UP(k)=E^\DO(k)$. Results for $E^{\UP\DO}(k)$ have been reported by
Wellein \etal\cite{WeRoFe96} and Wei{\ss}e \etal\cite{WeFeWeBi00,WeWeFe02}
However, in contrast to $A_\si(k,\en)$, $E^\UP(k)$ and $E^{\UP\DO}(k)$ do not
reveal the spectral weight of the corresponding QP's. Nevertheless, the
comparison with the spectra will yield valuable insight and serve as a test
of the CPT results. Moreover, a direct calculation of energy bands does not
suffer from the restricted energy resolution of CPT due to the use of a
smearing parameter [Eq.~(\ref{eq:Akw})].

Owing to the limitations regarding the number of phonon states, we shall only
show results for $\omb\geq1$. To be more specific, we consider two values of
the adiabatic ratio, namely $\omb=4$ and $\omb=1$. For $\omb=4$, the spectra
will turn out to be relatively simple, and we are able to study even strong
el-ph coupling. Consequently, we start with a discussion of the antiadiabatic
regime, and then move on to the more difficult case $\omb=1$.

\subsection{Antiadiabatic regime}\label{sec:nonadiabatic-regime}

\begin{figure*}[t]
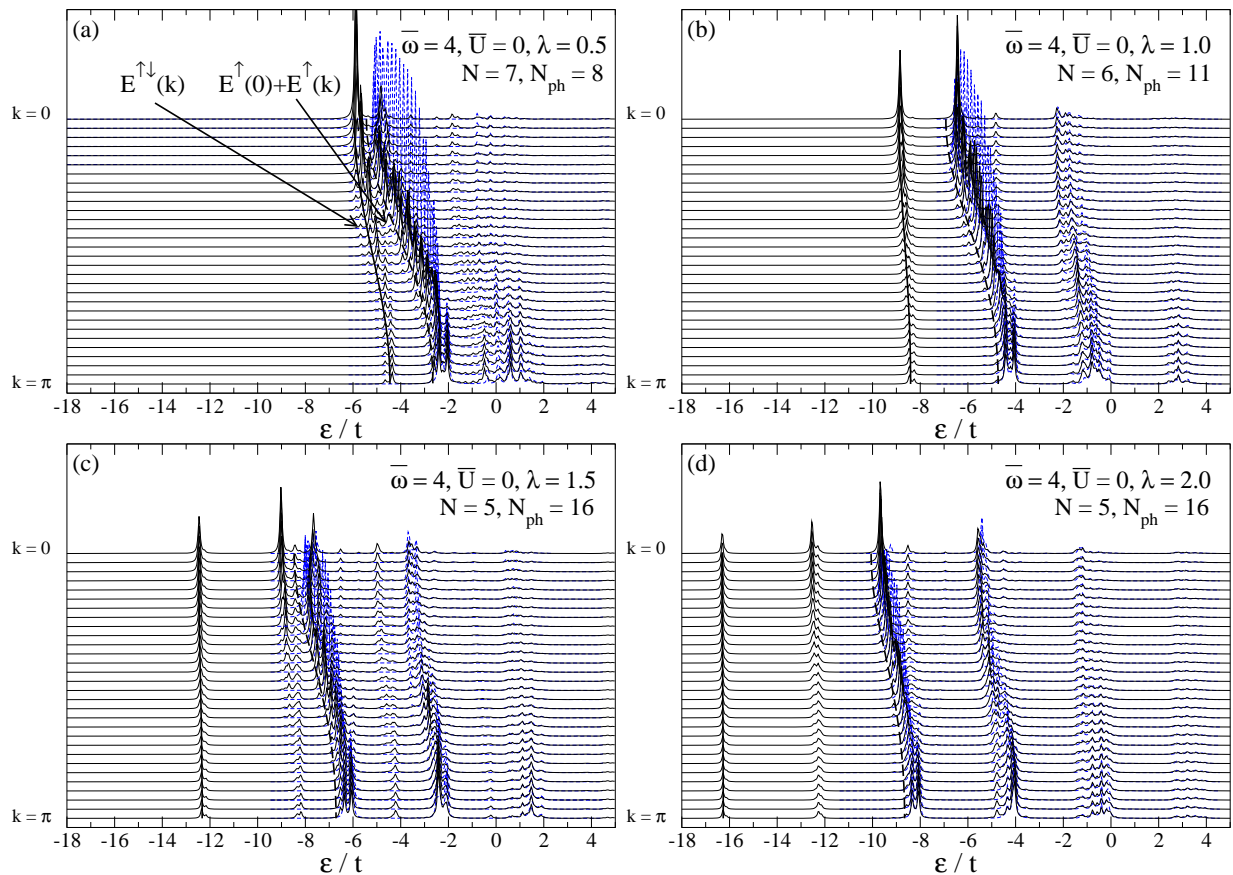

  \includegraphics[width=0.45\textwidth]{w4.0_U0.0_lambda0.5.eps}
  \includegraphics[width=0.45\textwidth]{w4.0_U0.0_lambda1.0.eps}\\
  \includegraphics[width=0.45\textwidth]{w4.0_U0.0_lambda1.5.eps}
  \includegraphics[width=0.45\textwidth]{w4.0_U0.0_lambda2.0.eps}
\caption{\label{fig:spectrum_k_w4.0_U0.0}
  (color online) Spectral functions $\sf{\UP}$ (solid lines) and $\sf{\DO}$
  (dashed lines) calculated with CPT for different values of the el-ph
  coupling $\lambda$, using $\eta=0.05t$ [see Eq.~(\ref{eq:Akw})].  All other
  parameters as indicated in the figures.  The truncation errors are
  $\Delta<$ (a) $5.3\times10^{-6}$, (b) $1.0\times10^{-5}$, (c)
  $2.4\times10^{-7}$, (d) $6.7\times10^{-6}$ (see text).  The vertical lines
  correspond to VDM results for the polaron and bipolaron band dispersions
  $E^\UP(0)+E^\UP(k)$ (dashed) and $E^{\UP\DO}(k)$ (solid),
  respectively.\cite{Sh03} }
\end{figure*}

\begin{figure}[t]
  \includegraphics[width=0.45\textwidth]{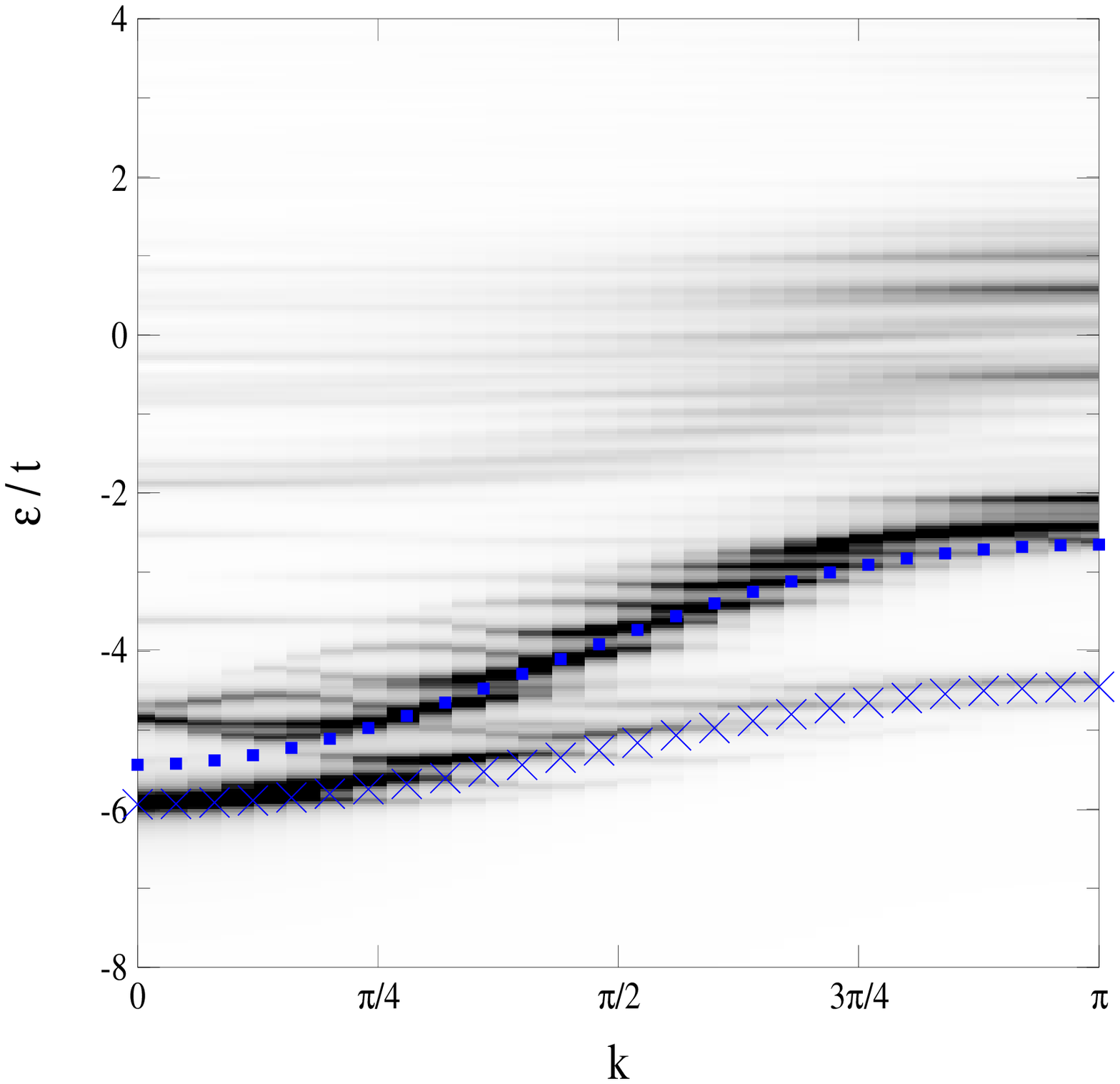}
\caption{\label{fig:greyscale}
  (color online) Density plot of the spectral function $\sf{\UP}$ for
  $\omb=4.0$, $\Ub=0$, and $\lambda=0.5$, as shown in
  Fig.~\ref{fig:spectrum_k_w4.0_U0.0}(a).  The symbols correspond to VDM
  results for $E^{\UP}(0)+E^\UP(k)$ (squares) and $E^{\UP\DO}(k)$ (crosses),
  respectively.\cite{Sh03}}
\end{figure}

In this section, we restrict the discussion to $\Ub=0$, while the influence
of Coulomb repulsion will be studied below.
Figure~\ref{fig:spectrum_k_w4.0_U0.0} shows the evolution of the one-particle
spectrum with increasing el-ph coupling.  Here and in subsequent figures,
solid lines represent results for $A_\UP$ and dashed lines correspond to
$A_\DO$.

For $\Ub=0$, two electrons of opposite spin always form a bipolaron state for
any $\lambda>0$. At weak coupling $\lambda=0.5$
[Fig.~\ref{fig:spectrum_k_w4.0_U0.0}(a)], $A_\UP$ exhibits two well visible
bands, as well as an incoherent part centered at $\en\approx0$. To understand
the nature of the coherent excitations, we have also included in
Fig.~\ref{fig:spectrum_k_w4.0_U0.0} the bipolaron band dispersion
$E^{\UP\DO}(k)$ (solid vertical line), calculated by the VDM.\cite{Sh03} The
latter fits well the low-energy band, with the minor deviations at
intermediate $k$---\ie, the splitting of the low-energy peak into several
small satellites---being finite-size effects, as has been verified by
calculations on smaller and larger clusters for a smaller number of phonon
states (not shown). A more detailed discussion of finite-size effects will be
given below for $\omb=1$.

Even at weak coupling $\lambda=0.5$, the bipolaron band already has a
relatively small width of $W'/W\approx0.37$ compared to the free-electron
value $W$. Moreover, the spectral weight of the lowest-energy peak, obtained
by integration over the CPT spectrum, decreases significantly from about
$0.68$ at $k=0$ to about $0.08$ at $k=\pi$. At the same time, the weight
contained in the incoherent part of the spectrum increases with increasing
$k$. This behavior is very similar to the single-electron
case.\cite{HoAivdL03}

We now turn our attention to the second, higher-lying band appearing in
Fig.~\ref{fig:spectrum_k_w4.0_U0.0}(a). From the general discussion in
Sec.~\ref{sec:holstein}, we expect that it corresponds to an excited state
with two polarons. We therefore compare it to the energy of two independent
polarons in an infinite system. Since $A_\UP$ describes the process of adding
an electron with momentum $k$ to the one-polaron ground state with energy
$E^\UP(0)$ [cf. Eq.~(\ref{eq:G_lanczos})], we show in
Fig.~\ref{fig:spectrum_k_w4.0_U0.0}(a) the band dispersion
$E^\UP(0)+E^\UP(k)$ (dashed vertical line). The comparison with the spectral
function yields a very good agreement at intermediate and large $k$, while
there are some discrepancies at small momenta. A density plot of $A_\UP$
(Fig.~\ref{fig:greyscale}) reveals more clearly that the two coherent bands
hybridize and repel each other near the point where they would be degenerate,
giving rise to an upper band with inversed dispersion at small $k$. The
situation is similar to the hybridization of the coherent and incoherent
parts in the one-electron case occurring for $|E^\UP(k) - E^\UP(0)|\sim\omb$
(see I). Of course, such effects are absent in the band dispersion of a
system with two independent polarons. Since the residual interaction between
two polarons vanishes in the limit $N\rightarrow\infty$, the hybridization
visible in the CPT spectrum may be attributed to finite-size effects. The
latter originate from the fact that within CPT, translational symmetry is
broken by treating inter- and intracluster hopping differently, and only
approximately restored afterward.

The spectral function $A_\DO$, also shown in
Fig.~\ref{fig:spectrum_k_w4.0_U0.0}(a), contains a coherent band at low
energies, and an incoherent part which is very similar to that of $A_\UP$.
Well away from $k=0$, the coherent peaks in $A_\DO$ follow closely the
polaron band in $A_\UP$. Thus the excited two-polaron state of the system
with two electrons of opposite spin is very similar to the ground
state of the system with two electrons of the same spin. Near $k=0$, the
spectral weight of the low-energy peak in $A_\DO$ is small ($\approx0.08$)
compared to the polaron peak in $A_\UP$ ($\approx0.2$). This is a result of
the fact that two polarons with the same spin cannot occupy
the same $k=0$ state. The picture changes at larger momenta, where both bands
have similar weight, although the sharp peaks in $A_\DO$ are higher than the
broadened features in $A^\UP$.

With increasing el-ph coupling, the low-energy bipolaron band becomes even
narrower until it is virtually flat at $\lambda=1.5$
[Fig.~\ref{fig:spectrum_k_w4.0_U0.0}(c)]. Here, the two conditions for a
small bipolaron (Sec.~\ref{sec:holstein}) are identical to $\lambda>0.5$.
Consequently, finite-size effects are very small in
Figs.~\ref{fig:spectrum_k_w4.0_U0.0}(b)\,--\,(d), as confirmed by the
excellent agreement of the CPT data with the results for $E^{\UP\DO}(k)$.
The reduction in bandwidth with increasing $\lambda$ is accompanied by a loss
of spectral weight. For $k=0$, the latter decreases from the value 0.68 at
$\lambda=0.5$ given above to about 0.10 at $\lambda=2.0$. Both the narrowing
and the loss of weight indicate a significant increase of the effective
bipolaron mass.

While the polaron band lies relatively close to the bipolaron band at
$\lambda=0.5$ [Fig.~\ref{fig:spectrum_k_w4.0_U0.0}(a)], the increase of the
coupling leads to a clear separation, and to a downward shift of both bands
proportional to the polaron binding energy $\Ep$. In the antiadiabatic
strong-coupling regime of Fig.~\ref{fig:spectrum_k_w4.0_U0.0}(d), the energy
gap between the two bands is well described by the atomic-limit value
$2\Ep=8t$. Similar to $\lambda=0.5$, the two-polaron band dispersion
$E^\UP(0)+E^\UP(k)$ agrees well with the polaron band in the spectra, with
some differences being visible near $k=0$. Interestingly, in
Fig.~\ref{fig:spectrum_k_w4.0_U0.0}(c), there is a mixing of the bipolaron
state with one phonon excited, which lies an energy $\om=4t$ above the lowest
band, and the two-polaron excitation.

The polaron band also narrows with increasing el-ph coupling (see also I).
However, the effect is much smaller than for the bipolaron band.
Additionally, the spectral weight of the $k=0$ polaron peak in $A_\UP$
increases from about 0.20 at $\lambda=0.5$ to about 0.32 at $\lambda=2.0$.
This may be explained by the fact that for weak coupling
[Fig.~\ref{fig:spectrum_k_w4.0_U0.0}(a)], some of the weight of the polaron
state is contained in the large low-energy feature. Calculations for a single
electron and the same parameters show that the spectral weight of the polaron
at $k=0$ decreases from about 0.86 ($\lambda=0.5$) to about 0.52
($\lambda=2.0$). Since the spectral weight is, to a very good approximation,
equal to the inverse of the effective mass of the Holstein
polaron,\cite{FeLoWe00} these results indicate that the polaron mass does not
increase at the same rate as the bipolaron mass with increasing coupling, as
reflected by the corresponding changes in bandwidth in
Fig.~\ref{fig:spectrum_k_w4.0_U0.0}. Finally, we also find a comparable
reduction of spectral weight for the two-polaron band in $A_\DO$ from about
0.08 ($\lambda=0.5$) to about 0.04 ($\lambda=2.0$) at $k=0$.

To conclude the discussion of the case $\omb=4$, we would like to underline
the enormous advantage of CPT in the strong-coupling regime. It permits us to
perform calculations on a very small cluster ($N=4$)---sufficient to obtain
well-converged results---but still yields the spectral function at any
desired $k$.

\subsection{Intermediate phonon frequency}\label{sec:interm-phon-freq}

\begin{figure*}[t]
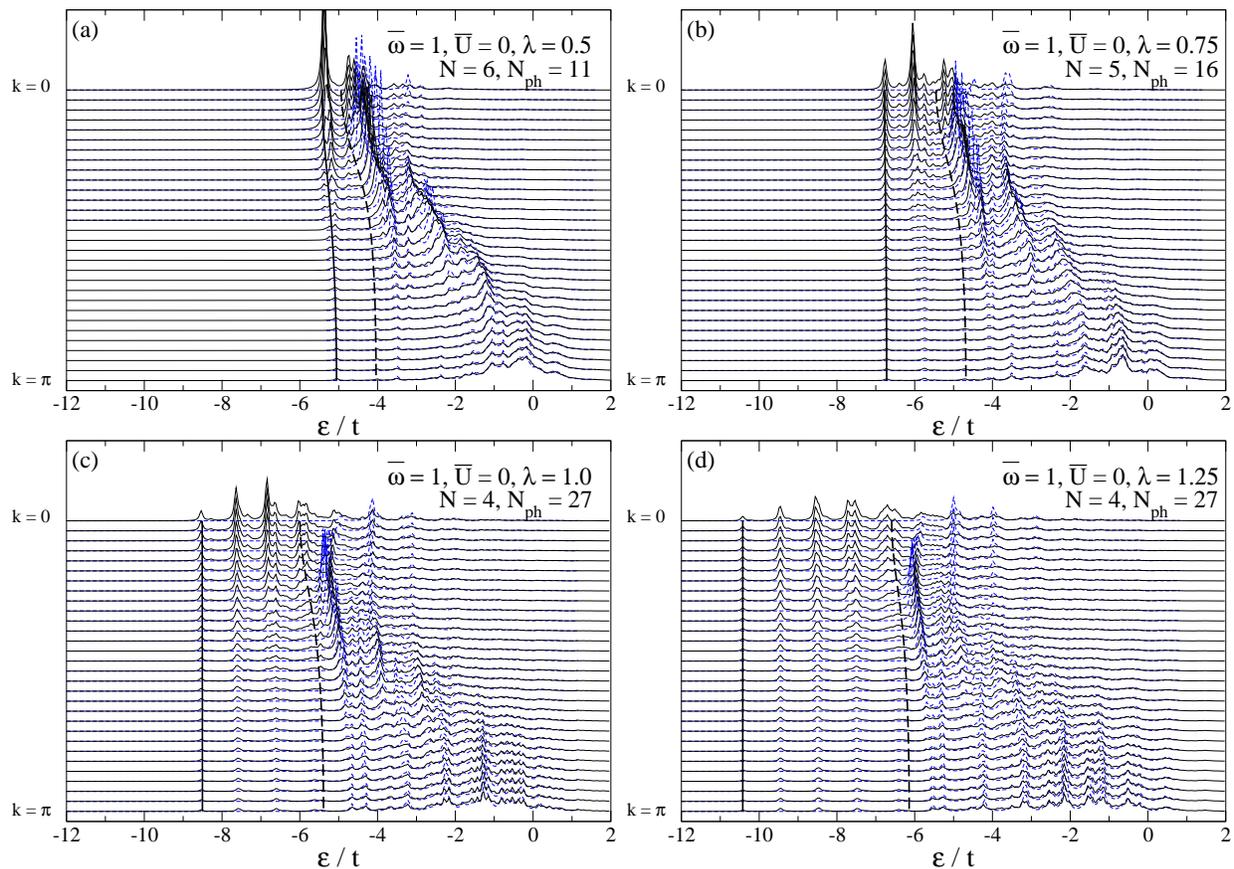

  \includegraphics[width=0.45\textwidth]{w1.0_U0.0_lambda0.5.eps}
  \includegraphics[width=0.45\textwidth]{w1.0_U0.0_lambda0.75.eps}\\
  \includegraphics[width=0.45\textwidth]{w1.0_U0.0_lambda1.0.eps}
  \includegraphics[width=0.45\textwidth]{w1.0_U0.0_lambda1.25.eps}
\caption{\label{fig:spectrum_k_w1.0_U0.0}
  (color online) Spectral functions $\sf{\UP}$ (solid lines) and $\sf{\DO}$
  (dashed lines) calculated with CPT for different values of the el-ph
  coupling $\lambda$, using $\eta=0.05t$. All other parameters as indicated
  in the figures.  The truncation errors are $\Delta<$ (a)
  $9.1\times10^{-5}$, (b) $9.0\times10^{-5}$, (c) $1.2\times10^{-7}$, (d)
  $2.4\times10^{-6}$.  The vertical lines correspond to variational
  diagonalization results for the polaron and bipolaron band dispersions
  $E^\UP(0)+E^\UP(k)$ (dashed) and $E^{\UP\DO}(k)$ (solid),
  respectively.\cite{Sh03} }
\end{figure*}

In the preceding section, we have investigated in detail the signatures of
polaron and bipolaron states in the one-particle spectrum for $\omb=4$. Owing
to the large energy of phonon excitations, most of the spectral weight
resides in the corresponding bands, allowing a fairly easy identification. We
now consider the case $\omb=1$, which turns out to be more difficult to study
numerically and to interpret. Nevertheless, work in the regime $\omb\leq1$ is
highly desirable to understand many interesting strongly correlated systems
such as, \eg, the manganites.  Although the latter are usually characterized
by $\om\ll t$, quantum effects are already visible for $\om=t$.  As a
consequence, previous
authors\cite{deRaLa86,RaTh92,ZhJeWh99,BoKaTr00,BoTr01,Mac04} have often
focused on this case, which is numerically much easier to tackle than the
region $\omb\ll1$. The case $\omb\ll1$ has been considered, \eg, by Wellein
\etal\cite{WeRoFe96} and Wei{\ss}e \etal\cite{WeFeWeBi00,WeWeFe02} While the
discussion for $\omb=4$ was restricted to $\Ub=0$, here we shall also take
into account a finite Coulomb repulsion.

\subsubsection{$\Ub=0$}

\begin{figure*}[t]
  \includegraphics[width=0.45\textwidth]{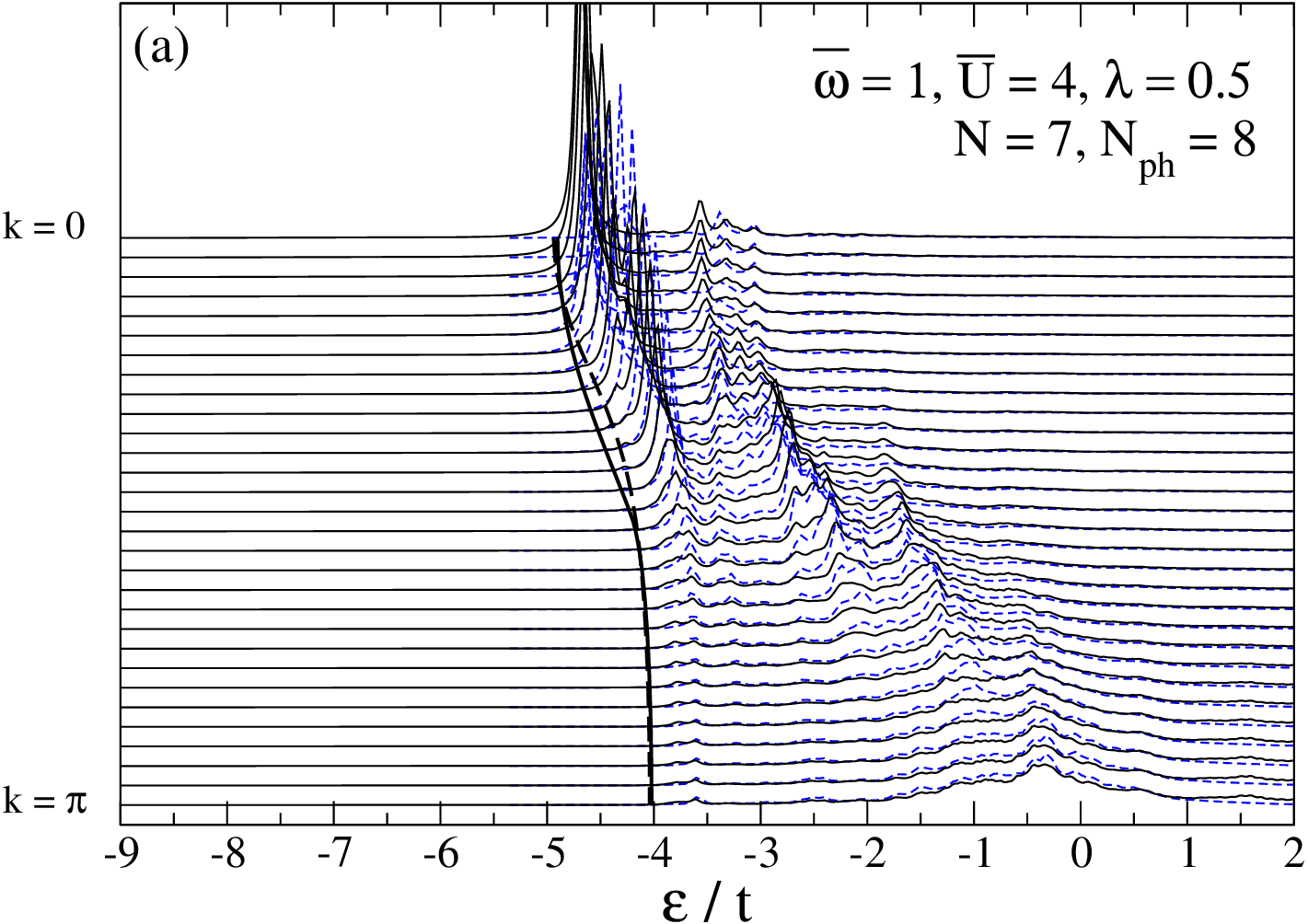}
  \includegraphics[width=0.45\textwidth]{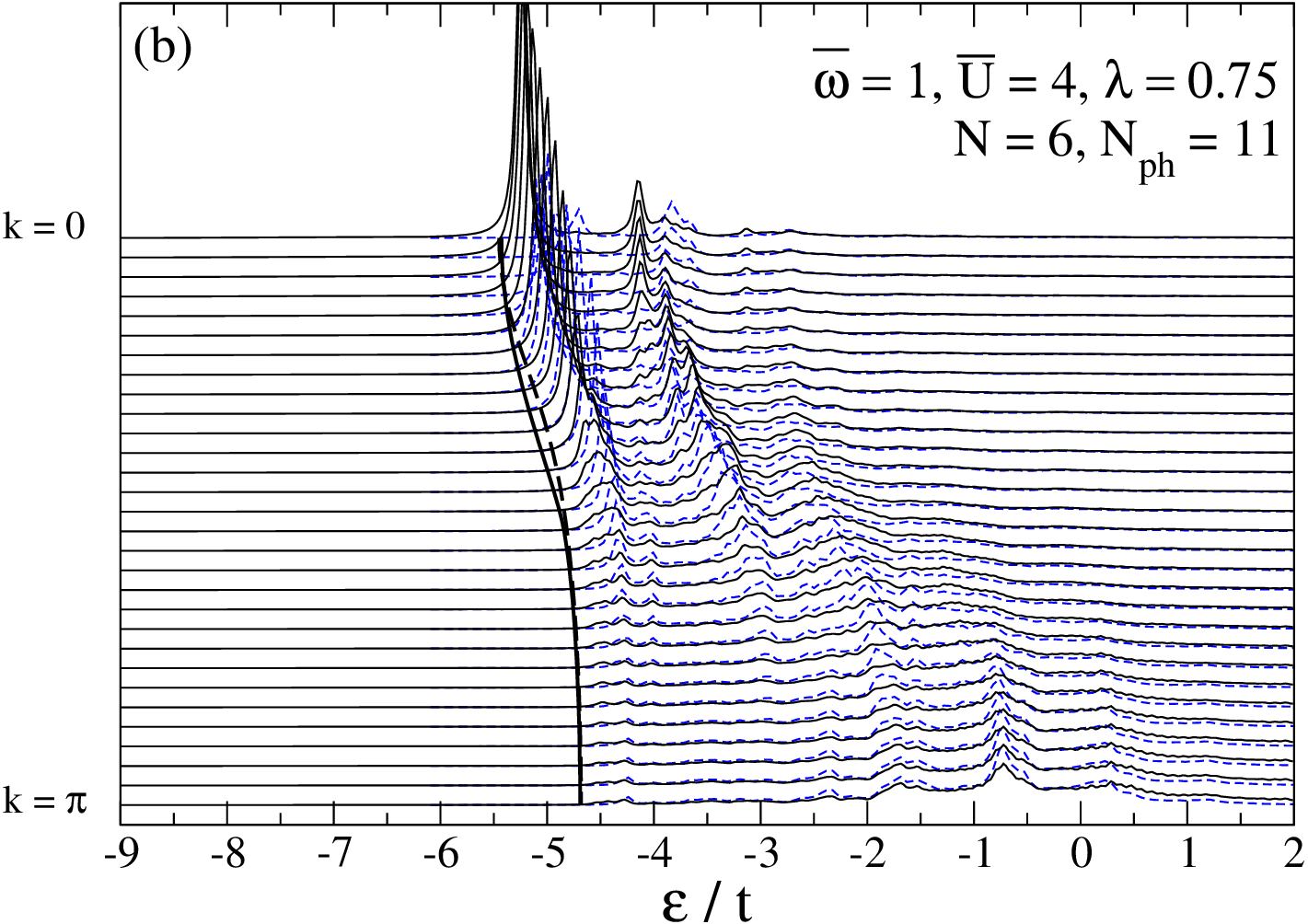}\\
  \includegraphics[width=0.45\textwidth]{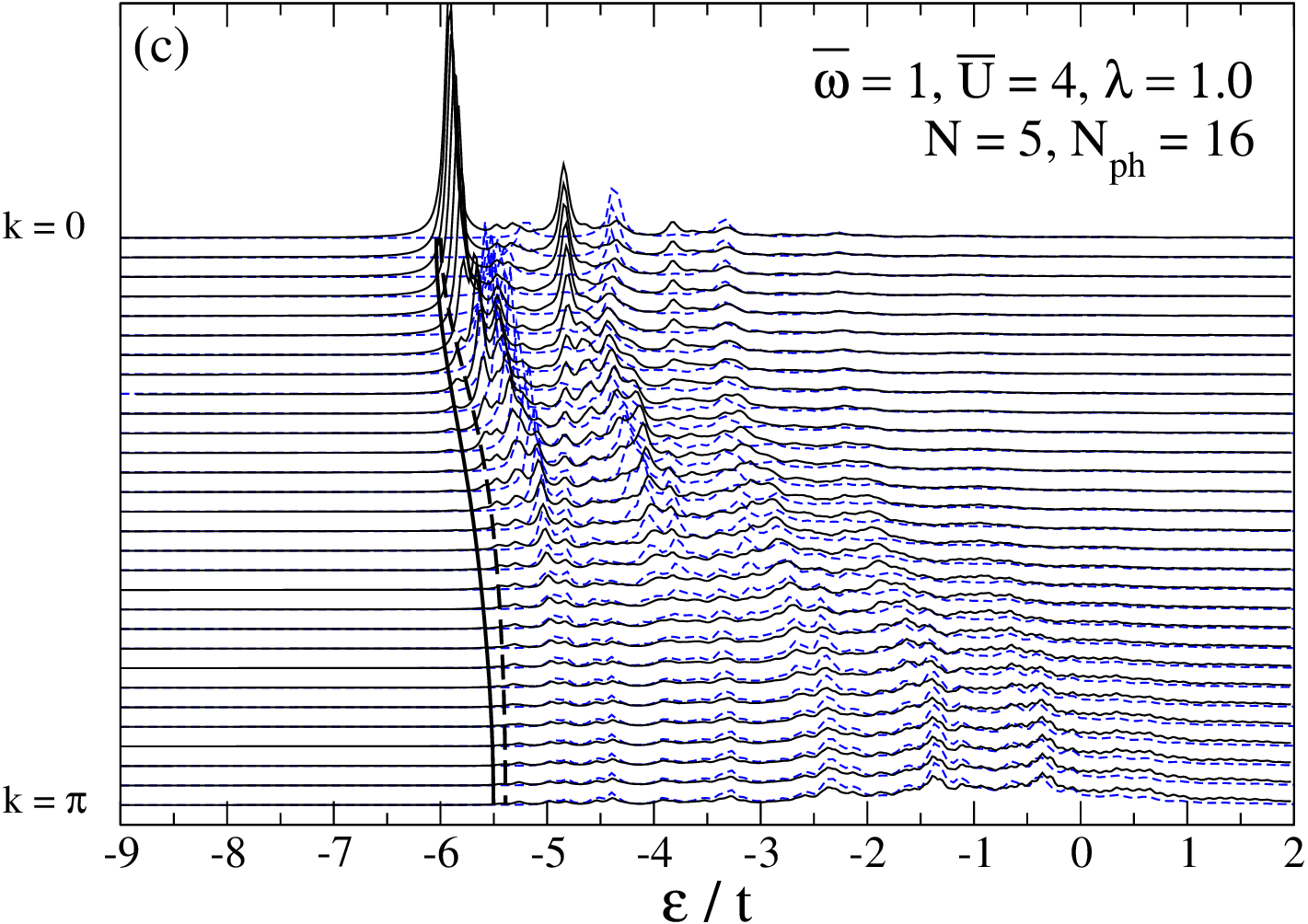}
  \includegraphics[width=0.45\textwidth]{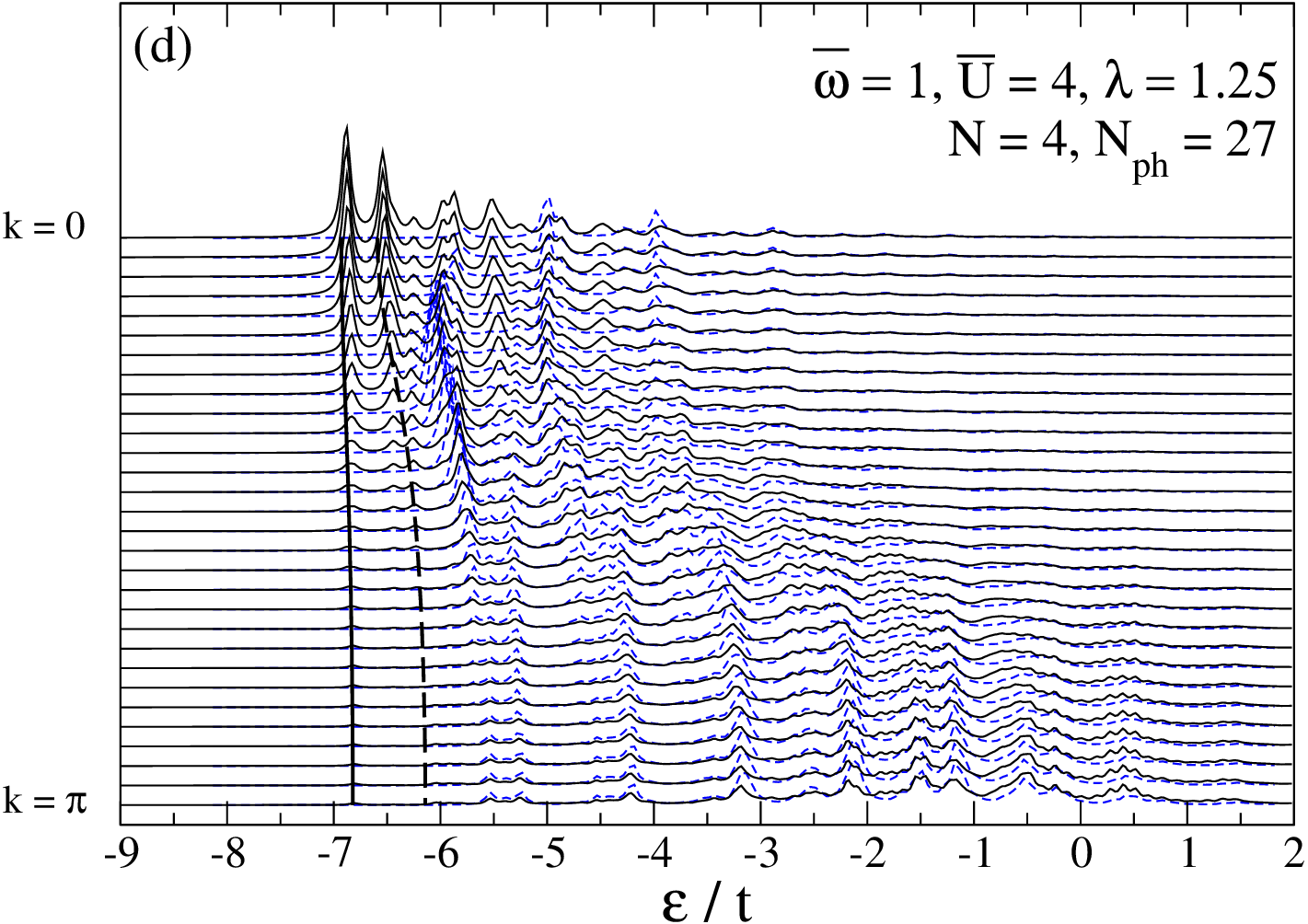}
\caption{\label{fig:spectrum_k_w1.0_U4.0}
  (color online) Spectral functions $\sf{\UP}$ (solid lines) and $\sf{\DO}$
  (dashed lines) calculated with CPT for different values of the el-ph
  coupling $\lambda$, using $\eta=0.05t$.  All other parameters as indicated
  in the figures.  The truncation errors are $\Delta<$ (a)
  $3.3\times10^{-5}$, (b) $2.0\times10^{-5}$, (c) $9.9\times10^{-6}$, (d)
  $6.7\times10^{-7}$. The vertical lines correspond to variational
  diagonalization results for the polaron and bipolaron band dispersions
  $E^\UP(0)+E^\UP(k)$ (dashed) and $E^{\UP\DO}(k)$ (solid),
  respectively.\cite{Sh03} }
\end{figure*}

Since converged results for $\omb=1$ require more phonon states than for
$\omb=4$, we have slightly reduced the cluster sizes in our calculations.
Consequently, finite-size effects are larger, as discussed below. Moreover,
we are not able to reach the strong-coupling regime but instead restrict the
range of $\lambda$ to 0.5\,--\,1.25.

Figure~\ref{fig:spectrum_k_w1.0_U0.0} contains the one-particle spectra for
$\Ub=0$. In principle, for $\lambda=0.5$, the results look quite similar to
Fig.~\ref{fig:spectrum_k_w1.0_U0.0}(a). However, the spectral weight of the
two coherent bands is much smaller, as a consequence of the increased
importance of incoherent excitations for $\omb=1$. In particular, the weight
of the latter is strongly enhanced at large $k$, so that the bands are no
longer easy to identify. Therefore, and because of the strong mixing of the
bands with coherent and incoherent excitations, it becomes difficult to
accurately determine the spectral weight by integration over the CPT spectra.

We see from Fig.~\ref{fig:spectrum_k_w1.0_U0.0} that the bipolaron bandwidth
is much smaller for $\omb=1$ ($W'/W\approx0.1$) than for $\omb=4$
[Fig.~\ref{fig:spectrum_k_w4.0_U0.0}(a)], despite the fact that the value of
$\lambda$ is the same in both cases. Hence, the effect of el-ph interaction
on the bipolaron mass is much more pronounced in or near the adiabatic regime
due to the larger mass of the oscillators.

In principle, the spectrum also contains coherent excited states which are
separated from the lowest-energy band by less than the phonon energy $\om$.
However, owing to the rather complex structure of the spectrum in the
two-electron case, they are difficult to distinguish from the other
contributions. A direct calculation of excited states in the Holstein model
with one electron has recently been presented by Bari\v{s}i\'{c}.\cite{Ba04}
Finally, the relation between $A_\DO$ and $A_\UP$ is very similar to
$\omb=4$.

As we increase the el-ph coupling, the bipolaron dispersion collapses to an
extremely narrow band [Fig.~\ref{fig:spectrum_k_w1.0_U0.0}(b)]. This cross
over is again associated with a significant loss of spectral weight. At
$k=0$, for example, we find a reduction from about 0.50 at $\lambda=0.5$ to
about 0.14 at $\lambda=0.75$.  Increasing $\lambda$ further to 1.25, we
finally arrive at a bipolaron band with $W'/W\approx10^{-4}$ and a spectral
weight of less than 0.03 at $k=0$.  Similar to
Fig.~\ref{fig:spectrum_k_w4.0_U0.0}(d), the spectrum displays several bands
equally spaced by $\om$, which belong to states with one or more phonons
excited. Moreover, the polaron and bipolaron bands are well separated, and
the incoherent contributions dominate at large $k$.

The agreement between the bipolaron band dispersion and $E^{\UP\DO}(k)$ in
Fig.~\ref{fig:spectrum_k_w1.0_U0.0} is again very good. Similar to $\omb=4$,
the condition for a small bipolaron is given by $\lambda>0.5$, so that CPT
yields very accurate results. In contrast, the two-polaron energy
$E^\UP(0)+E^\UP(k)$ fits less well to the corresponding bands in the spectral
function. We attribute this difference to the antiadiabatic regime
(Fig.~\ref{fig:spectrum_k_w4.0_U0.0}) to the stronger retardation effects for
$\omb=1$. As a consequence, the polaron state is more extended below the
small-polaron cross over occurring at $\lambda=1$ (see, \eg, I), leading to a
stronger residual interaction on a finite cluster, which also manifests
itself in the CPT results. In contrast, for $\omb=4$, the lattice
distortions around the electrons are very localized, and the two polarons are
almost independent. Above $\lambda=1$, \ie, in the small-polaron regime, the
two-polaron dispersion for $\omb=1$ again follows closely a two-polaron-like
feature in the spectrum [Figs.~\ref{fig:spectrum_k_w1.0_U0.0}(c) and~(d)].

\subsubsection{$\Ub=4$}

\begin{figure*}[t]
  \includegraphics[width=0.5\textwidth]{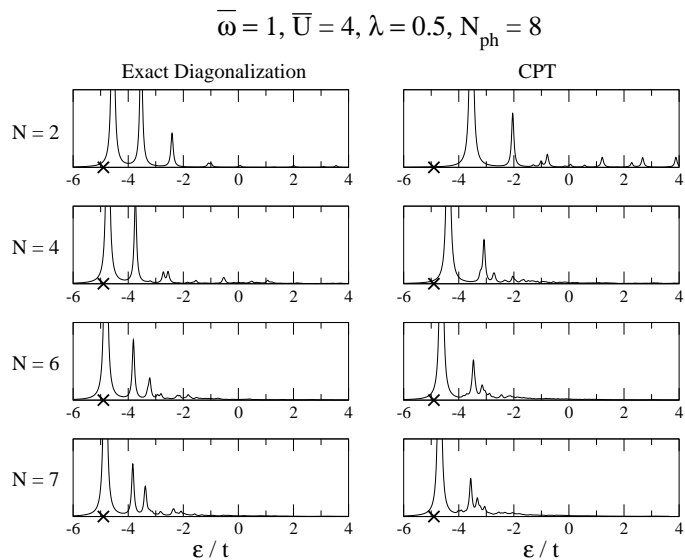}
\caption{\label{fig:size_U4}
  Comparison of the spectral function $A_{\UP}(0,\en)$ calculated with ED and
  CPT, respectively, for different clusters sizes $N$, using $\eta=0.05t$.
  The crosses correspond to the VDM result 
  for the bipolaron energy $E^{\UP\DO}(0)$.\cite{Sh03}
}
\end{figure*}

So far, we have only presented results for $\Ub=0$, for which a bipolaron
state is always favored. However, in materials such as the cuprates or the
manganites, strong local correlations hinder the carriers from forming onsite
bipolarons even for strong el-ph coupling. To model such effects, we
therefore consider here a finite value of the el-el repulsion $\Ub=4$.

In the case of two electrons with opposite spin, the Lanczos results for the
cluster Green function converge faster as a function of $\Nph$ for $\Ub>0$
compared to $\Ub=0$ as a result of the reduced effective el-el interaction.
This is fortunate, since it allows us to use slightly larger clusters,
thereby partly compensating for the increased finite-size effects due to the
spatially more extended ground state in the weak-coupling regime.

From the general discussion in Sec.~\ref{sec:holstein}, we expect the ground
state to consist of two weakly bound polarons for $2\Ep<U$, and a cross over
to a bipolaron state at a critical value of the el-ph interaction $\lambda$.
In the antiadiabatic limit, the latter is determined by $2\Ep=U$ (\ie,
$\lambda=1$ for the case considered here) for weak coupling, and by $4\Ep=U$
for strong coupling.

In Fig.~\ref{fig:spectrum_k_w1.0_U4.0}, we present the results for the
spectral function, again for $\lambda=0.5$\,--\,1.25. For weak coupling
$\lambda=0.5$ [Fig.~\ref{fig:spectrum_k_w1.0_U4.0}(a)], the most striking
difference to the $\Ub=0$ case discussed above is the fact that there appears
only one band at low energies. Together with the incoherent contributions,
and taking into account the doubling of the number of carriers leading to a
shift of energies, the spectrum bears a close resemblance to that of a single
polaron with the same parameters (Fig.~3 of I).  This is also underlined by
the polaron and bipolaron band dispersions shown in
Fig.~\ref{fig:spectrum_k_w1.0_U4.0}(a), which are almost identical throughout
the Brillouin zone, and lie just below the corresponding band in the spectral
function. In particular, the band displays the typical flattening at large
$k$, where the low-energy excitations have mostly phononic character.
Furthermore, owing to the finite onsite repulsion, the low-energy band in
$A_\DO$ is very similar to that in $A_\UP$ since, for finite $\Ub$ and weak
el-ph coupling, the singlet ground state consists of two weakly bound
polarons. Consequently, the singlet and triplet state have comparable
energies, although the spectral weight in $A_\DO$ is again very small near
$k=0$.

For $\lambda=0.75$ [Fig.~\ref{fig:spectrum_k_w1.0_U4.0}(b)], the ground state
of the system is still given by two polarons, and the spectrum is almost
indistinguishable from $\lambda=0.5$. In the present case, the condition for
the existence of an intersite bipolaron is expected to lie between the weak-
and strong coupling results $U<2\Ep$ and $U<4\Ep$.\cite{BoKaTr00} However,
owing to its small binding energy, the intersite state is difficult to
distinguish from the two-polaron state in the spectral function.

At $\lambda=1.0$ [Fig.~\ref{fig:spectrum_k_w1.0_U4.0}(c)], the band in
$A_\UP$ begins to split. Although the energy difference between the polaron
and bipolaron band dispersions is still relatively small near $k=0$, an
excitation gap clearly emerges at larger $k$. Finally, at $\lambda=1.25$, two
distinct bands with similar spectral weight have formed, which agree very
well with $E^{\UP\DO}(k)$ and $E^\UP(0)+E^\UP(k)$, respectively.
Interestingly, the band in the triplet spectral function $A_\DO$ lies
noticeably higher than the polaron band in $A_\UP$. Thus, for the parameters
considered here, two polarons of opposite spin can lower their total energy
by occupying the same lattice site, which is just the mechanism behind
bipolaron formation.

The abovementioned discrepancies between the bipolaron band dispersion
$E^{\UP\DO}(k)$ obtained by Shawish and the band in $A_\UP$ are a result of
finite-size effects in the CPT calculations. The latter become smaller with
increasing coupling $\lambda$ together with the size of the bipolaron, and
for $\lambda=1.25$ we find a very good agreement
[Fig.~\ref{fig:spectrum_k_w1.0_U4.0}(d)]. To illustrate this point, we
compare in Fig.~\ref{fig:size_U4} the spectral function $A_\UP$ at $k=0$,
$\lambda=0.5$, and for different cluster sizes $N$, calculated using ED with
periodic boundary conditions (left column) and CPT (right column),
respectively. The results reveal that for weak coupling and intermediate
$\Ub$, ED is superior to CPT concerning the convergence of the peak positions
with respect to system size.  This is not surprising as CPT is based on a
strong-coupling expansion in the hopping term.\cite{SePePL00} Here, the el-el
and el-ph interaction are both of about the same magnitude as the hopping, so
that the method does not work as well as for $\Ub=0$.

For $\Ub>0$, finite-size effects in both, CPT and ED, are larger due to the
extended bipolaron state which exists for weak coupling.  Similar to the
one-electron case discussed in I, deviations from the exact results due to
the finite cluster size are usually smallest for $k=0$, while they become
larger with increasing $k$. Although in Fig.~\ref{fig:size_U4} the positions
of the peaks in the CPT spectral function are slightly less accurate than in
the case of ED, the weights of the excited states resemble more closely the
results in the thermodynamic limit.

Finally, for $\Ub>4$, the cross over to a small bipolaron occurs at even
larger values of $\lambda$. Apart from the change of the critical coupling,
the physics is not altered significantly. Therefore, we have restricted our
discussion of the spectral function to $\Ub\leq4$, but some results for the
bipolaron band dispersion at $\Ub=8$ will be presented below.

\subsubsection{Bipolaron band dispersion}

The bipolaron band dispersion $E^{\UP\DO}(k)$ has been calculated before by
Wellein \etal\cite{WeRoFe96} and Wei{\ss}e \etal\cite{WeFeWeBi00,WeWeFe02}
for small phonon frequencies $\omb=0.4$ and $\omb=0.5$, respectively.
Remarkably, for parameters $\Ub>0$ and $\lambda>0$ such that the effective
interaction $U_\text{eff}=0$ [Eq.~(\ref{eq:Ueff})], they found a
renormalized, free-particle dispersion relation.\cite{WeFeWeBi00,WeWeFe02} In
this section, we wish to extend these considerations to the case $\omb=1$,
and to infinite systems.  While the narrowing effect due to el-ph interaction
has been discussed above, here we focus on the form of the band.

\begin{figure}[t]
  \includegraphics[width=0.45\textwidth]{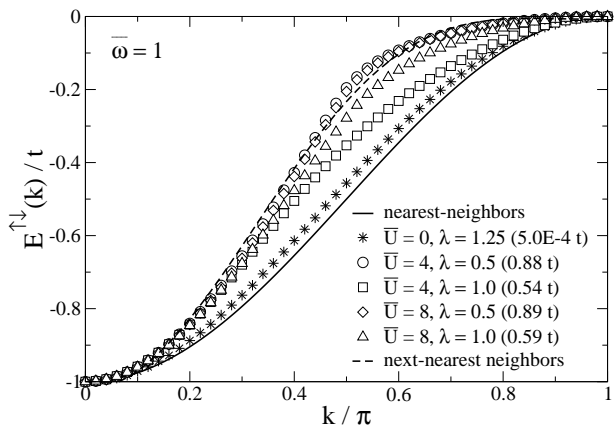}
\caption{\label{fig:dispersion}
  Bipolaron dispersion $E^{\UP\DO}(k)$ as a function of the wavevector
  $k$.\cite{Sh03} Also shown is the bare tight-binding dispersion for
  nearest-neighbor hopping, and a fit to the results for $\Ub=4$, $\lambda=0.5$
  using a dispersion for nearest-- and next-nearest neighbor hopping (see
  text). All curves have been scaled to the interval $[-1,0]$, with the
  actual bandwidths given in the legend.}
\end{figure}

Owing to the limited energy resolution and finite-size effects in the CPT
results shown above, we use the more accurate data from the VDM. In
Fig.~\ref{fig:dispersion}, we show Shawish's\cite{Sh03} results for the
bipolaron energy as a function of $k$, for different values of $\Ub$ and
$\lambda$.  To permit a direct comparison, we have scaled all curves to the
interval $[-1,0]$, with the actual bandwidths given in the legend.

We begin with the regime of a strongly bound small bipolaron. To this end we
consider the case $\Ub=0$ and $\lambda=1.25$. The corresponding band
resembles quite closely a cosine dispersion, with some deviations being
visible around $k=\pi/2$. A different behavior is found for finite $\Ub=4$,
as well as weak coupling $\lambda=0.5$. For these parameters, which favor a
ground state with two polarons [see Fig.~\ref{fig:spectrum_k_w1.0_U4.0}(c)],
the form of the band is remarkably different from a simple tight-binding
dispersion for nearest-neighbor hopping. This is still true for $\lambda=1$,
although a trend towards a cosine dispersion is visible. For even larger
$\Ub=8$, the noncosine-like form persists even for $\lambda=1$.

It is worth mentioning the great similarity of the results for $\Ub=4$ and
$\Ub=8$ in the weak-coupling regime, which follows from the fact that once
the small (onsite) bipolaron state is energetically unfavorable for the two
electrons due to the Coulomb repulsion, a further increase of the latter has
very little effect. On top of that, the intersite bipolaron state which
exists for $\Ub>0$ has a very small binding energy, so that the band
dispersion is almost the same as that of two polarons.

To identify the origin of the deviations from a free-electron band, we also
included in Fig.~\ref{fig:dispersion} a fit of a free-electron model with
nearest and next-nearest neighbor hopping to the band for $\Ub=4$ and
$\lambda=0.5$, which yields an amplitude $t'\approx0.6t$ for two-site hopping
processes.  As proposed by Wellein \etal,\cite{WeRoFe96} the importance of
long-range hopping for the band dispersion of a single polaron may be due to
a residual polaron-phonon interaction, with the phonons and the polaron
residing on different sites. Since we find substantial deviations of the
bipolaron band from a cosine dispersion only in the regime of two weakly
bound polarons, it stands to reason to assume the same underlying mechanism.

Finally, we would like to comment on the fact that despite $U_\text{eff}=0$
for $\Ub=4$ and $\lambda=1$ in Fig.~\ref{fig:dispersion}, we do not have a
simple cosine band, in contrast to the findings of Wellein
\etal\cite{WeRoFe96} and Wei{\ss}e \etal,\cite{WeWeFe02} which have have been
attributed to the formation of an intersite bipolaron.\cite{WeWeFe02} In
contrast, here we observe noncosine-like behavior even in the regime where an
intersite state exists. These differences are expected to be a result of the
larger value of the phonon frequency (here $\omb=1$, while $\omb=0.4$ and
$0.5$ in Refs.~\onlinecite{WeRoFe96} and~\onlinecite{WeWeFe02},
respectively), leading to a noticeable reduction of retardation effects.
Moreover, since the critical coupling $\lambda_\text{c}$ decreases as
$\omb\rightarrow0$, the bipolaron is more strongly bound in the work of
Refs.~\onlinecite{WeRoFe96,WeWeFe02}, thereby suppressing the abovementioned
nonlocal phonon-polaron interaction. Further work along these lines is highly
desirable to understand the dependence of the bipolaron band dispersion on
the phonon frequency in the regime $\omb\leq1$.

\section{\label{sec:summary}Conclusions}

We have presented a detailed study of the one-electron spectral function of
the Holstein-Hubbard model with two electrons of either the same or opposite
spin. The method employed here is cluster perturbation theory together with
the Lanczos method, which represents a versatile and fast approach.

As a function of the electron-phonon and electron-electron interaction
strength, polaron and bipolaron states manifest themselves as quasiparticle
bands, and results have been compared to accurate data for the bipolaron
energy dispersion. For weak coupling and/or intermediate to strong Hubbard
repulsion, finite-size effects are visible, but are much smaller than in
previous work restricted to small clusters. The major advantage of the
present method is that the spectrum can be obtained at any point in
$\bk$-space, even when using clusters with only a few lattice sites for which
enough phonon states can be kept in the calculation. This has allowed us to
investigate, for the first time, the dispersion and the spectral weight of
the quasiparticle features throughout the Brillouin zone.  The results and
their dependence on the model parameters have been discussed, and a perfect
agreement has been found with the physical picture of the Holstein-Hubbard
bipolaron emerging from previous work. A comparison of the bipolaron
dispersion with a simple tight-binding band has revealed an important
contribution from next-nearest-neighbor hopping processes in the regime of a
weakly bound state.

Finally, the adiabatic regime of small phonon frequencies, characteristic of
many real materials, remains an interesting and demanding open issue for
future work.

\begin{acknowledgments}
  
  This work has been supported by the Austrian Science Fund (FWF), project
  No.~P15834.  M.~H. and M.~A. are grateful to DOC (Doctoral Scholarship
  Program of the Austrian Academy of Sciences). We are indebted to
  David~M.~Eagles for stimulating correspondence, and to Samir El Shawish for
  useful discussions, as well as for providing us with previously unpublished
  data. Finally, we would like to thank Holger Fehske for making valuable
  comments on the manuscript.
  
\end{acknowledgments}



\end{document}